\documentclass{aastex}
\usepackage{spr-astr-addons}
\usepackage{url}

\RequirePackage{color}

\usepackage{graphicx}
\usepackage{amsmath}	
\usepackage[dvipsnames]{xcolor}  
\usepackage{amssymb}
\usepackage{newtxtext,newtxmath}

\usepackage[colorlinks]{hyperref}

\hypersetup{
  colorlinks=true,        
  linkcolor=blue,         
  citecolor=cyan,         
}

\begin{document}

\title{Neutron and quark stars: constraining the parameters for simple EoS using the GW170817.}
\shorttitle{Neutron and quark stars}
\shortauthors{ Arroyo-Chav\'ez et al.}
\author{Griselda Arroyo-Ch\'avez \altaffilmark{1,2}}
\email{g.arroyo@irya.unam.mx}
\author{Alejandro Cruz-Osorio\altaffilmark{3,4}}
\email{osorio@itp.uni-frankfurt.de }
\author{F. D. Lora-Clavijo\altaffilmark{5}}
\email{fadulora@uis.edu.co}
\author{Cuauhtemoc Campuzano Vargas \altaffilmark{2}}
\and 
\author{Luis Alejandro Garc\'ia Mora\altaffilmark{6}} 
\email{306229969@ciencias.unam.mx}

\altaffiltext{1}{Instituto de Radioastronom\'ia y Astrof\'isica, UNAM, Campus Morelia, A.P. 3-72, C.P. 58089, M\'exico.}
\altaffiltext{2}{Facultad de F\'isica, Universidad Veracruzana, 91000, Xalapa Veracruz, M\'exico.}
\altaffiltext{3}{Universitat de Val\'encia, Dr.  Moliner 50, 46100, Burjassot (Val\'encia), Spain.} 
\altaffiltext{4}{Institut f{\"u}r Theoretische Physik, Max-von-Laue-Stra{\ss}e 1, 60438 Frankfurt, Germany.}
\altaffiltext{5}{Grupo de Investigaci\'on en Relatividad y Gravitaci\'on Escuela de F\'isica, Universidad Industrial de Santander A. A. 678, Bucaramanga 680002, Colombia.}
 \altaffiltext{6}{Instituto de F\'{\i}sica y Matem\'{a}ticas, Universidad Michoacana de San Nicol\'as de Hidalgo.  Edificio C3, Cd. Universitaria, 58040 Morelia, Michoac\'{a}n,  M\'{e}xico.}

\begin{abstract}
It is well known that the equation of state (EoS) of compact objects like neutron and quark stars 
is not determined despite there are several sophisticated models to describe it. From the 
electromagnetic observations, summarized in \cite{Lattimer01}, and the recent observation of 
gravitational waves from binary neutron star inspiral GW170817 \cite{Abbott2017_etal} and 
GW190425 \cite{Abbott2019}, it is  possible to make an estimation of the range of masses and 
so constraint the mass of the neutron and quark stars, determining not only the best approximation 
for the EoS, but which kind of stars we would be observing. In this paper we explore several 
configurations of neutron stars assuming  a simple polytropic equation of state, using a single 
layer model without crust. In particular, when the EoS depends on the mass rest density, 
$p=K \rho_{0}^{\Gamma}$,  and when it depends on the energy density $p=K \rho^{\Gamma}$, 
considerable differences in the mass-radius relationships are found. On the other hand, we also 
explore quark stars models using the MIT bag EoS for different values of the vacuum energy density $B$. 
\end{abstract}

\keywords{sample article; }

\section{Introduction}
\label{sec:intro}
The new astronomy and astrophysics of multi-mes\-sen\-gers were born from the event GW170817, 
in which the gravitational and electromagnetic radiation coming from the collision of a binary 
neutron star system was measured for the first time \cite{Abbott2017_etal, Abbott2017b}. The 
masses of neutron stars responsible for this strong emission of gravitational waves have been 
estimated in the range $1.17  \ M_{\odot}- 1.6 \ M_{\odot}$ with a total mass of the system of 
$2.74^{+0.04}_{-0.01} M_{\odot}$ \cite{Abbott2017_etal}. In new detection during the third observing 
run (O3) of the LIGO-Virgo  detectors in the event GW190425, the estimated masses are in the 
range $1.45 \ M_{\odot}-1.88 \ M_{\odot}$ for low spinning neutron stars with total mass 
$2.3^{+0.1}_{-0.1} \ M_{\odot}$  \cite{Abbott2019}. In the last decades, the study of neutron stars has 
become one of the branches of relativistic astrophysics of more interest in the scientific community, since these objects are extremely dense and there is uncertainty about the behave of matter inside. 
In order to understand a little more about the behaviour of these ultracompact objects, 
many general relativistic numerical simulations have been carried out to extract the gravitational 
waveform coming from the collision of neutron stars for different equations of state  
\cite{Andersson:2009yt,Baiotti2016,tsokaros2017,East2016}, 
\cite[see also a recent review][]{Paschalidis2017b}. Several EoS have been used to model a neutron 
star, making a comparative analysis of its mass, radius and binding energy \cite{Lattimer01}. Recently in Most et al. (2018) and Rezzolla et al. (2018), the maximum mass and radius of 
these objects has been constrained to 
$2.01^{+0.04}_{-0.04} M_{\odot}<M<2.16 ^{+0.17}_{-0.5}M_{\odot}$ and 
$12~{\rm km}<R<13.45~{\rm km}$, respectively. The constraint was obtained after studying 
millions of equilibrium neutron stars with different equations of state.

Since the observation of gravitational waves, the event GW170817, the join of observation 
of gravitational waves and numerical simulation will become an important tool 
for astronomy and astrophysics. One of the challenges is finding a state equation for 
neutron stars which reproduces the waveforms observed. 
In this sense, a widely accepted candidate as EoS is the quark star \cite{Alcock86, Itoh70}, which 
is composed by stable strange quark matter (general Witten's conjecture) 
\cite{Bodmer1971,Witten84,Farhi1984}.  Moreover, the quark stars satisfy the tidal deformability 
estimated from the observation of the gravitational wave event GW170817  \citep{Lai2017b}. 
The usual EoS to describe a fluid composed by mixed strange quark matter under nuclear forces 
is the MIT bag model \cite{Farhi1984}.  Recently, general relativistic simulations were performed 
in order to study the axisymmetric and triaxial solutions of uniformly rotating quark stars 
\cite{Zhou2018,Zhou2018a}.

In this work, we solve the Tolman-Oppenheimer-Volkoff (TOV) equations with a single layer 
polytropic EoS and the MIT bag model to construct models of non-rotating neutron and quarks 
stars respectively, in order to determine which set of parameters of the EoS describe the masses 
and radii reported from the observations \citep{Lattimer01,Abbott2017_etal}. The paper is 
organized as follow:  In section \ref{sec:TOV}, we briefly describe the TOV equations in 1D 
spherical symmetry as well as the numerical details used to solve such system of equations. 
Moreover, we add a brief description of the EoS used to model the neutron and quark stars. We 
show our numerical result for neutron and quark stars in section \ref{sec:Results}, and finally 
some final comments in section \ref{sec:Sum}. Hereafter, we use the Einstein convention on 
sums over repeated indices and use geometrized units where $G = c = 1$.

\section{ Tolman-Oppenhaimer-Volkoff equations} 
\label{sec:TOV}

In this section, we built neutron and quarks stars models with different central densities. The 
models are carried out by solving numerically the Tolman-Oppenheimer-Volkoff (TOV) equations 
for a spherically symmetric static spacetime described by the line element
\begin{eqnarray}
ds^{2} =-{\alpha}^{2}d{t}^ {2} 
	           + \frac{d{r}^{2}}{1-\frac{2m(r)}{r}} 
	           + {r}^{2}\big(d{\theta}^{2} 
	           + {sin}^{2}\theta d{\phi}^{2}\big), \label{eq:metric}
\end{eqnarray}
\noindent being $m(r)$ the gravitational mass function inside the radius $r$, and 
$\alpha= \alpha(r)$ the lapse function associated with 3+1 formalism in general relativity. 
The matter used in the model  is a perfect fluid described by the energy-momentum tensor 
${T}^{\mu \nu} = \left(\rho_{0} + \rho_{0}\epsilon + p\right){u}^{\mu}{u}^{\nu} + p{g}^{\mu \nu}$, 
where $\rho_{0}$ is the baryon rest mass density, $\epsilon$ is the specific internal energy density, 
$p$ is the fluid pressure,  ${u}^{\mu}$ are the components of the 4-velocity and  $g_{\mu \nu}$ 
are the  components of the 4-metric  (\ref{eq:metric}). It is worth mentioning that $\rho_{0}$ and 
$\epsilon$ are related through the following relation $\rho=\rho_{0}(1+\epsilon)$ called energy 
density.  

Assuming the fluid inside the star is in hydrostatic equilibrium, the TOV equations are given as a 
system of ordinary differential equations for $m$, $p$ and $\alpha$
\begin{eqnarray}
\frac{dm}{dr} &=&  4 \pi r^{2} \rho, \label{eq:mass} \\
\frac{dp}{dr} &=& -(\rho +p)\frac{m + 4 \pi r^{3} p}{r(r - 2m)},  \label{eq:press} \\
\frac{1}{\alpha} \frac{d\alpha}{dr} &=& \frac{ m  +4 \pi r^{3} p }{ r(r -2rm)} \label{eq:alp}. 
\end{eqnarray}
To integrate the system of equations, it is necessary to introduce an equation of state to close the system. Particularly in this work, we consider two EoS, the first to model neutron stars 
which consist of a single polytropic EoS and the second to model quark star corresponding to
 the MIT bag model EoS.  
Both equations of state will be described in the next sections. The TOV equations become 
singular at $r=0$ that is avoided performing a Taylor expansion around this point and assuming 
the contour conditions $m(r=0)=m''(r=0)=m'''(r=0)=0$. Moreover, we use a guess constant value 
for $\alpha=0.5$ and impose the conditions $\alpha(r_{max})=1/a(r_{max})$, where 
$a(r)^2=g_{rr}=1/(1-2m/r)$ \cite[see for more details]{Guzman2012}. These conditions satisfy 
that outside of the star the solution is given by the Schwarzschild space-time. On the other hand, 
the surface of the star is defined where the rest mass density is equal to the atmosphere density 
$\rho_{\rm atm}=1\times 10^{-10}$ in geometric units.

The numerical solutions of the TOV equations were carried out by using the CAFE code 
\cite{Lora2015} \cite[see also][]{Lora2013,Cruz2012,Cruz2016,Lora2015219,Cruz-Osorio:2017epa} 
with a third order total variation diminishing Runke-Kutta integrator \cite{Shu88} in 1D spherical 
coordinates. The domain extends from $r_{min}=0$ to $r_{max}$, which is chosen depending on 
the model. In all simulations we use a uniform spatial grid with spatial resolution $\Delta r = 0.06$.

\subsection{ Polytropic Equation of State}
\label{subsec:NSEoS}

The polytropic EoS, which was introduced for the first time by 
\cite{tooper_1965_afs}, corresponds to a relation between the pressure and the rest mass density 
profiles $\rho_0$, {\it case 1}. However, there is another expression of EoS in a way the pressure 
depends on the energy density $\rho$ instead of the rest mass density \cite{Tooper64}, {\it case 2}. 

{\it Case 1:}  The polytropic equation of state has been used to describe a completely degenerate 
gas in Newtonian theory and general relativity. Traditionally, in this EoS, the pressure 
is written as a function of the rest mass density as follows
\begin{equation}
p = K \rho_0^{\Gamma} = K \rho_0^{1+1/n}, \label{eq:EoS1}
\end{equation}
where $K$, $\Gamma$ and $n$ are usually called the polytropic constant, polytropic exponent, 
and polytropic index, respectively. In this case, the energy density is related to the pressure by 
\begin{eqnarray}
\rho=\big( \frac{p}{K}\big)^{1/\Gamma} + \frac{p}{(\Gamma-1)},
\label{eq:polirest} 
\end{eqnarray}
where it has been assumed that the specific internal energy satisfies the ideal gas equation of state 
$\rho_0 \epsilon = p/(\Gamma - 1)$.

{\it Case 2:} The second possibility we have considered in this work assumes that the rest mass 
density is replaced by the energy density in the polytropic EoS, as follows
\begin{eqnarray}
\rho=\left( \frac{p}{K}\right)^{1/\Gamma},
\label{eq:politotal}
\end{eqnarray}
\noindent where the rest mass density is computed through the definition 
$\rho=\rho_{0}(1+ \epsilon)$ for comparison purposes. 

In this model the density and pressure profiles in neutron stars usually decay from center 
to the surface in adiabatic fashion, where in the entropy gradients are neglected  $dS = 0$, 
i.e, the specific entropy is constant. 

\subsection{MIT bag Model for Quark Stars}
\label{subsec:MITEoS}

Typically, quark stars are modelled with an equation of state based on  MIT bag-model of quark 
matter \cite{chodos1974}  \citep[see also][]{limousin2005, Zhou2018}, which satisfies the weak 
interaction and neutral charge condition. Neglecting the strange quark mass the EoS is given by 
the simple formula
\begin{eqnarray}
p =\frac{1}{3}(\rho - 4B),
\label{eq:MIT}
\end{eqnarray}
where $\rho$ and $B$ are the energy and vacuum energy densities of the bag that contain the 
confined quarks with three different colors: up, down and strange. From lattice QCD calculations 
is known that a phase transition from quarks (confined to nucleons) to  free quarks occurs before 
a density of $6\rho_{\rm{nuc}}$ is reached, being $\rho_{\rm{nuc}} = 2.3 \times 10^{14} ~g/cm^{3}$ 
the nuclear saturation density  \cite{Paschalidis2017b}. For stable strange quark matter the vacuum 
energy density values ranges from $57~MeV/fm^{3}$ to $92 ~MeV/fm^{3}$ 
\cite{Schmitt2010} \footnote{$1~~ MeV/fm^{3} = 1.6022 \times 10^{33}~dyn/cm^{2}$}, 
nevertheless, a more recently work reports slightly different ranges 
$58.926~MeV/fm^{3} < B < 91.5~ MeV/fm^{3}$ \cite{Paschalidis2017b}. It is worth mentioning 
that there are more sophisticated equations of state, which involve interacting quarks 
\cite{Flores:2017kte} or more complex structures, which involve anisotropic quark stars with an 
interacting quark EoS \cite{Becerra-Vergara:2019uzm}.

\section{Results}
\label{sec:Results}

\subsection{Neutron stars} 
\begin{figure}[h!]
\center
\includegraphics[width=1.0\columnwidth]{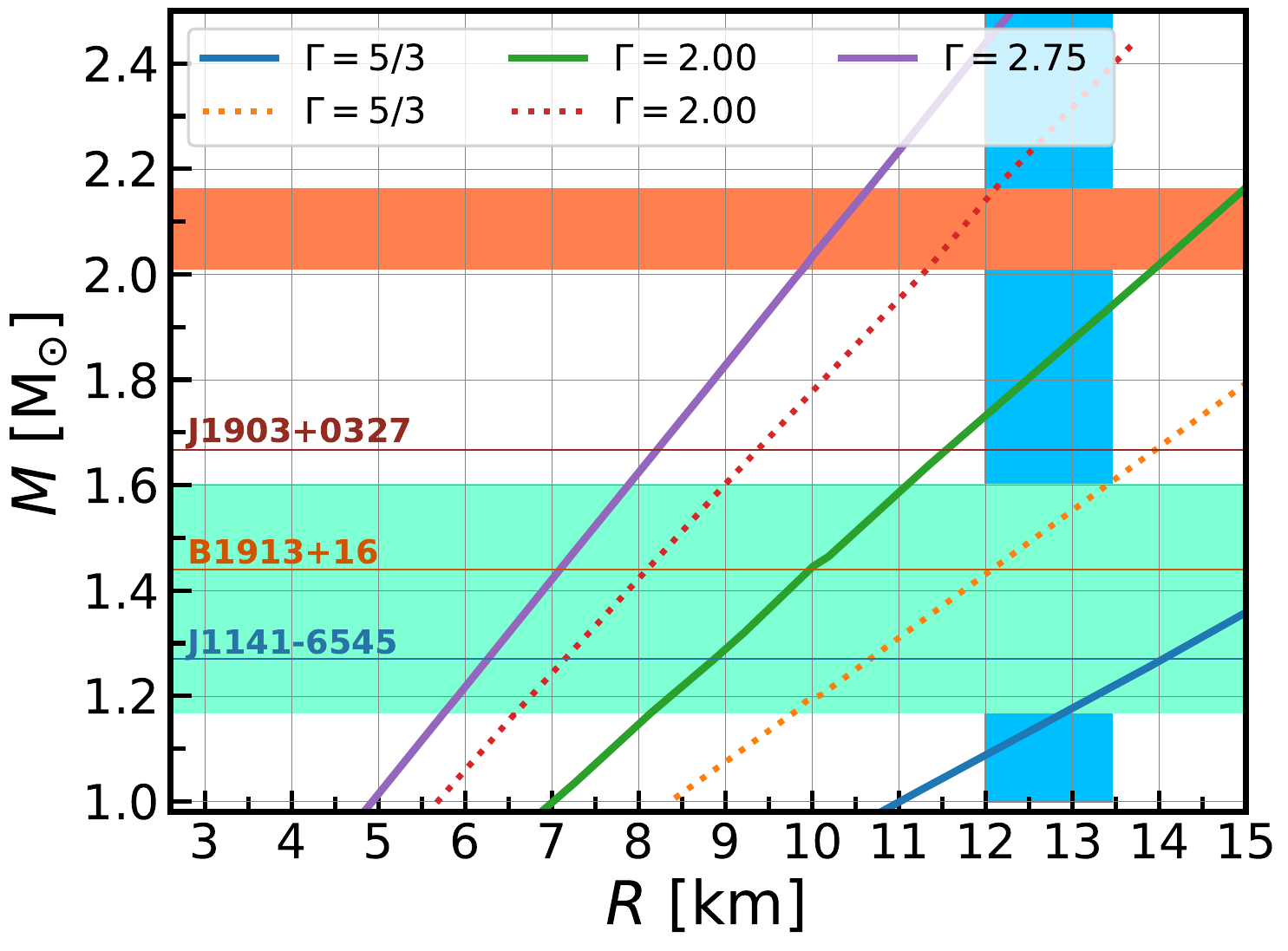} 
\caption{We show the compactness of neutron stars, i.e, the maximum mass versus 
radius of the several configurations of neutron stars. The solid lines shows the models with EoS 
defined with rest mass density \eqref{eq:polirest} and the dotted lines corresponds to the EoS defined 
using the energy density \eqref{eq:politotal}. We include, in horizontal lines three masses estimated from observed 
pulsars \cite{Lattimer01}. The light green cover range of estimated masses of individual neutron
star from gravitational waves emission \cite[]{Abbott2017_etal}. Also shown are in orange and 
blue shaded regions the recent  maximum mass and radii constrictions \cite{Rezzolla2017,Most2018}. 
\label{fig:neutronstars}}
\end{figure}

The neutron stars models are constructed by solving the system of equations (\ref{eq:mass}), 
(\ref{eq:press}), (\ref{eq:alp}) coupled with the state equations (\ref{eq:polirest}) and (\ref{eq:politotal}) 
with different central densities. For build up our physical understanding of the neutron stars, 
in this work we explore different values of the adiabatic index  $\Gamma=1.1$, $~4/3$, $~ 5/3$, $~1.181$, $1.87$, $~1.93$, $~ 2.0$, $~2.02$, $~2.05$, $~2.15$, $~2.24$, $~2.40$, $~2.75$, 
$~5.0$. The results are depicted in Figure \ref{fig:neutronstars}, where we show 
the total mass of the star as a function of the radius - the called compactness of the star - for EoS with the adiabatic index  $\Gamma$ and constant $K$ such that the sound-speed in the star is less than the speed of light. We found that for equation \eqref{eq:polirest} the acceptable adiabatic index range is $\Gamma \in [5/3, 2.75]$ (see the solid lines) while for the EoS given en equation \eqref{eq:politotal} the range is $\Gamma \in [5/3, 2.0]$ (see the dotted lines), respectively.

In this figure,  green  shaded region we also depict the range of non-rotating neutron stars masses ($1.17<M<1.6~ M_{\odot}$) estimated from the gravitational wave event GW170817 \cite{Abbott2017_etal},  
the orange region corresponds to the maximum mass interval obtained from the recent constriction 
of neutron stars \cite{Rezzolla2017} and finally the blue fringe is the acceptable radius for neutron stars 
\cite{Most2018}. We also include some particular pulsars like J1141-6545, B1913+16, and J1903+0327 
with masses $M=1.27 \pm 0.01 M_{\odot}$, $1.4398 \pm 0.0002 M_{\odot}$ and $1.667 \pm 0.021 M_{\odot}$, respectively, choosing the pulsars with smallest error bars which are reported in \cite{Lattimer01}. 
It is worth mentioning that all the lines were constructed by using several values of the parameter 
$K$ for each equation of state. We found that the neutron stars modelled with EoS (\ref{eq:politotal}) 
are more compact than the ones where the polytropic EoS is applied directly over the rest mass 
density. This difference is more noticeable for smaller values of $\Gamma$ than for higher ones. 
For instance, for a given value of $K$, $\Gamma = 5/3$ and a neutron star mass of 
$1.2~M_{\odot}$, the radius difference is about $\sim 3~{\rm km}$, being the model with EoS 
(\ref{eq:politotal}) the one with the smallest radius, while for $\Gamma = 5$   such difference is 
about $\sim 0.5~{\rm km}$. From this figure, we can conclude that when the adiabatic index 
increases the radius mass relation for neutron stars tends to be the same for both equations of 
state. 

\begin{figure}[h!]
\center
\includegraphics[width=0.85\columnwidth]{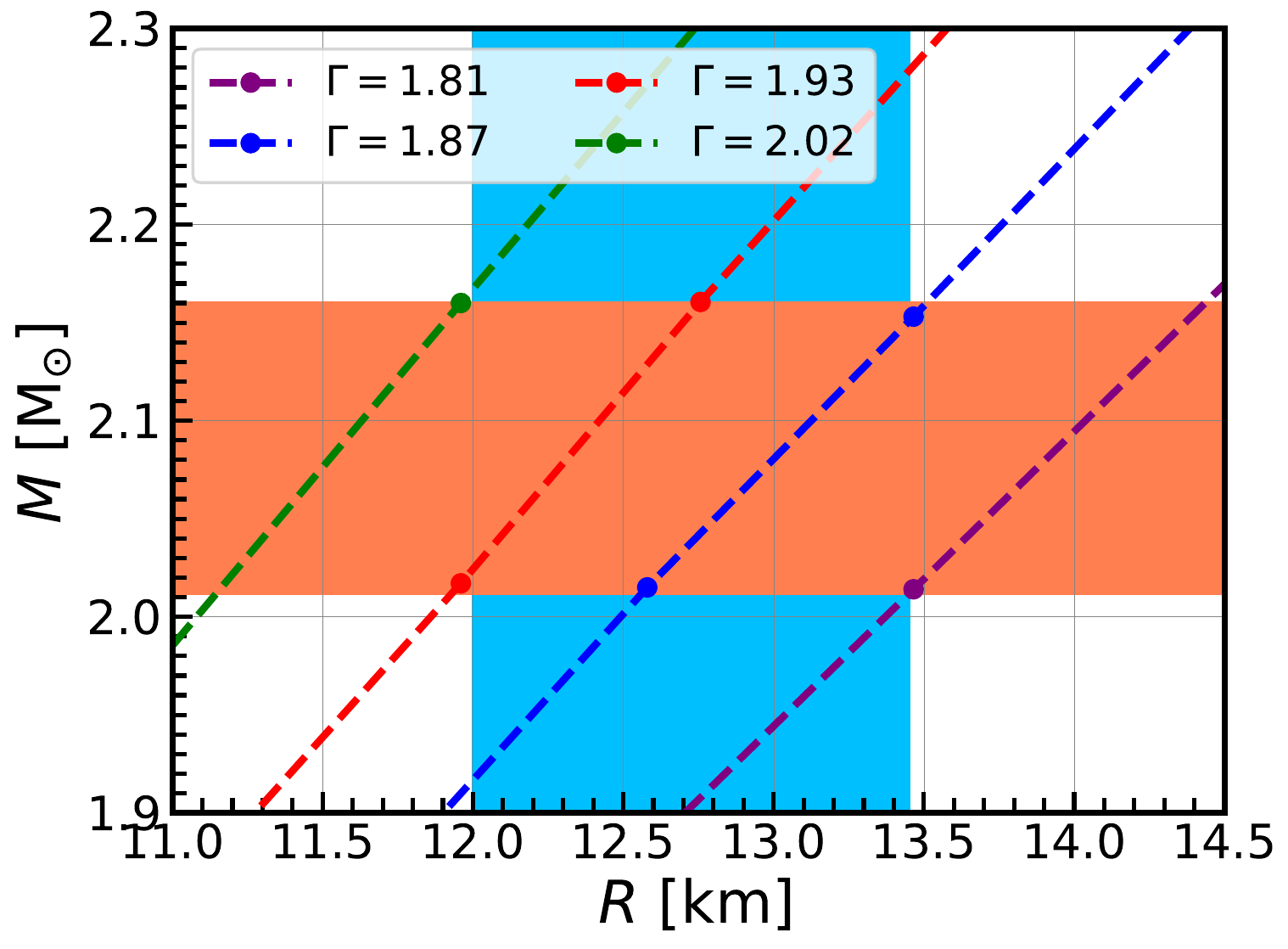} \\
\includegraphics[width=0.85\columnwidth]{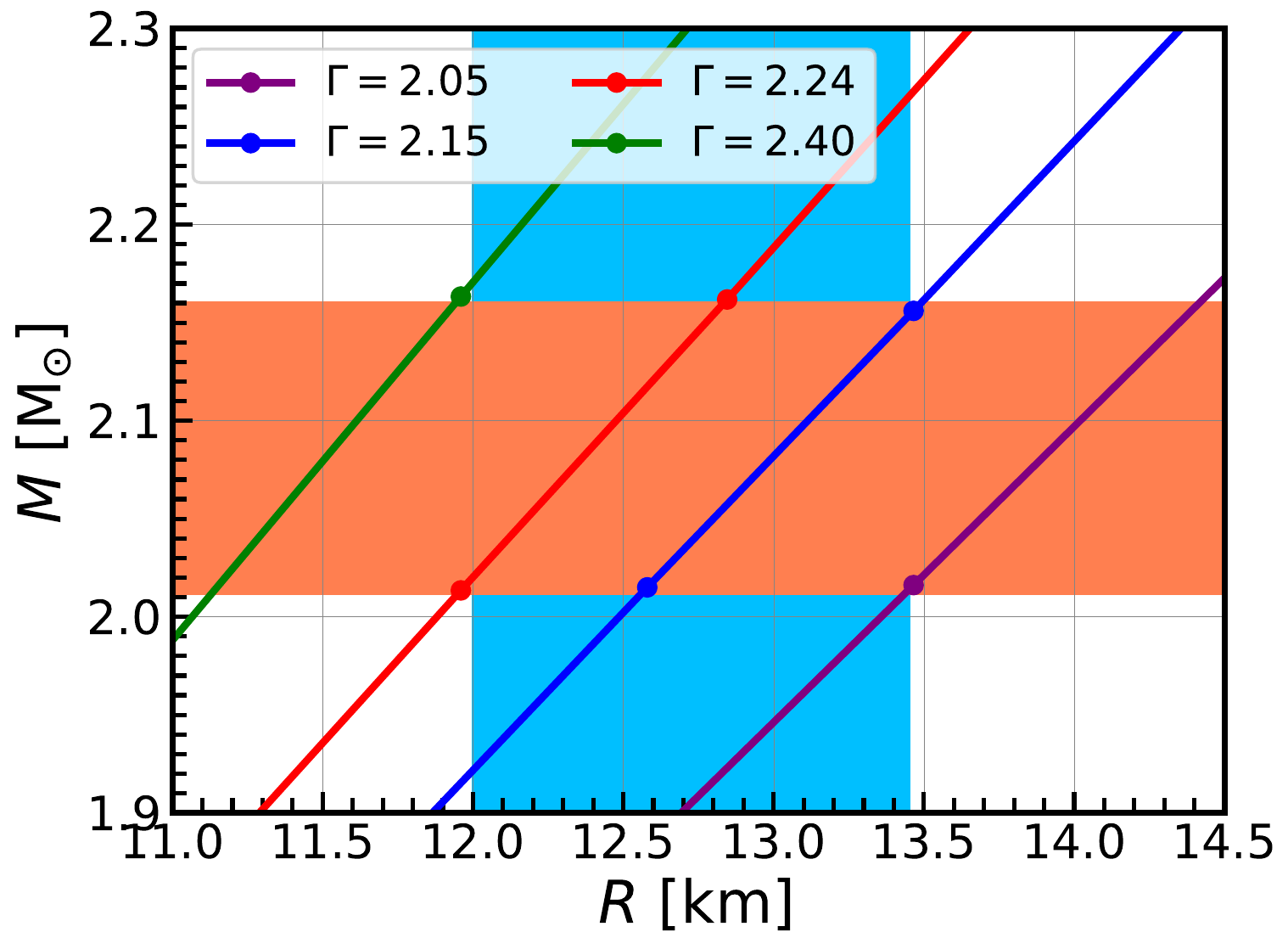}
\caption{Close-up view of the Figure \ref{fig:neutronstars} to highlight maximum mass and the 
corresponding radii. We show the fine-tuned values for $\Gamma$ and $K$ such that we cover 
mostly the range of masses and radii reported in \cite{Rezzolla2017, Most2018} for both cases,
using the rest mass density in EoS (top panel) and when the pressure is written as function of the energy 
density (bottom panel). See Table \ref{tab:par}, for corresponding $\Gamma$ and $K$ to the represented by the dots.}
\label{fig:fit}
\end{figure}
Now by  assuming that the maximum mass range and radius are $2.01<M<2.16~M_{\odot}$ and 
$12<R<13.45 ~{\rm km}$, respectively \cite{Rezzolla2017}, we realize a fit on the adiabatic index 
and constant $K$ in order to see what values of these parameters, for both EoS, are necessary to 
satisfy this constriction. In Figure \ref{fig:fit}, we show the corresponding values of $\Gamma$ that 
satisfies the constraint over the masses and radius for neutron stars reported in \cite{Rezzolla2017}. 
Specifically, we found for the {\it case 1}  values that range from  $\Gamma=2.05$ to $\Gamma=2.40$, while for the {\it case 2} these values range from  $\Gamma=1.81$ to $\Gamma=2.02$. The respective values of constant $K$, adiabatic index, effective temperature, density and pressure are reported in table  \ref{tab:par}. 

\begin{figure}[h!]
\center
\includegraphics[width=1.0\columnwidth]{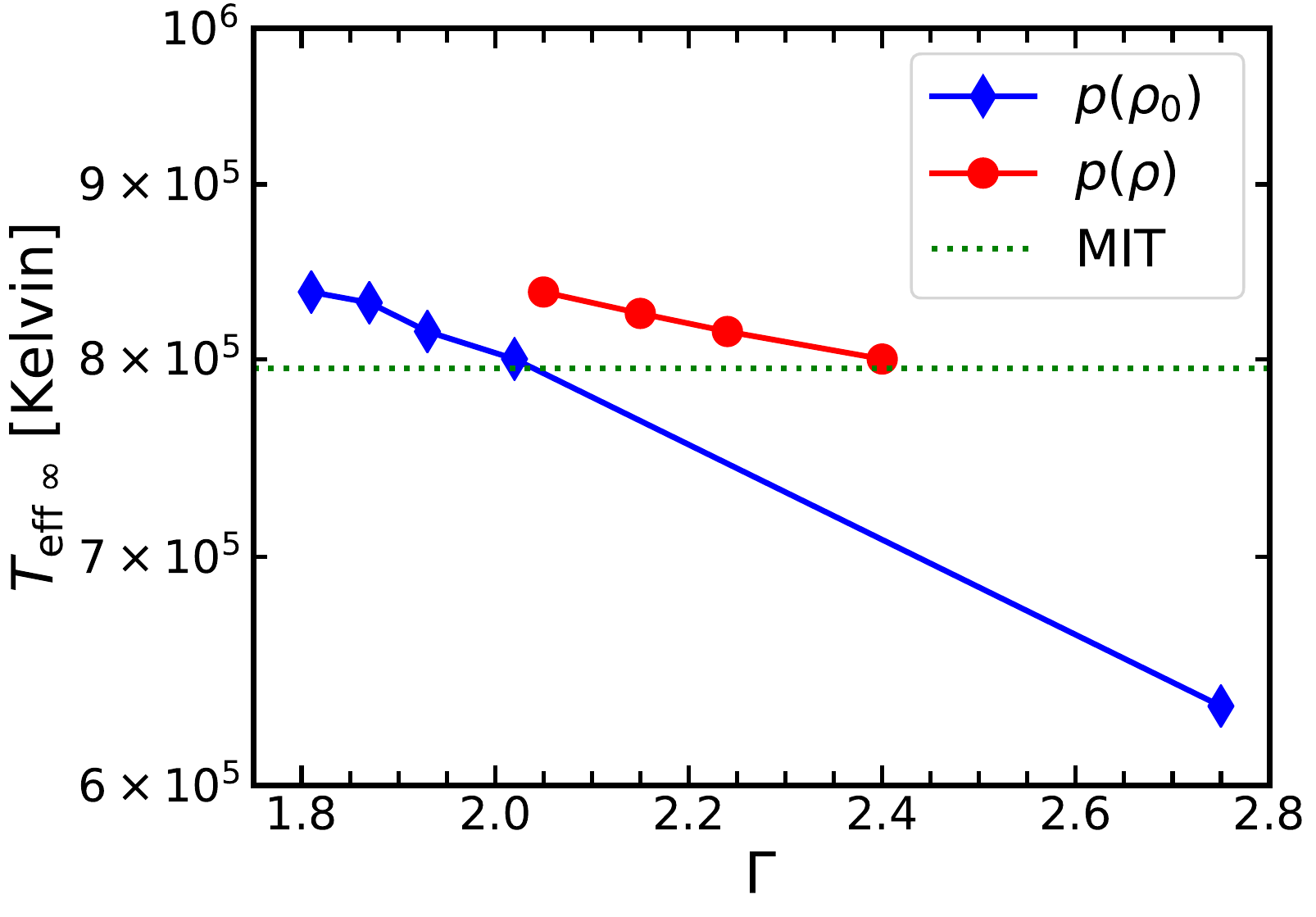}
\caption{Average of the effective temperature expressed either in Kelvin degrees measured at infinity versus the adiabatic index $\Gamma$.  Shown with blue and red lines for EoS given in equations \eqref{eq:polirest} and \eqref{eq:politotal} and in dotted line we depict the MIT Eos case. The effective temperature at the star surface is $T_{\rm eff}=10^6 \ {\rm K^{o}}$ \cite{Yakovlev2004}. See also Table \ref{tab:par} where we report the $T_{\rm eff \ \infty}$  for models reported in Figure \ref{fig:fit}.}
\label{fig:teff}
\end{figure}

Additionally to the global aspect of neutron stars, {\it i.e}, the compactness, we compute the effective temperature from neutron stars surface.  Is well known that the thermal emission from the surface of neutron stars are estimated from pulsars  giving the effective temperatures in the range $3 \times 10^{5}  \ K^{o} - 10^6 \ K^{o}$ \cite{Lattimer04,Page2004}.  The effective temperature definition come up from theoretical cooling calculation trough the Stefan-Boltzmann law $L_{\gamma}:=4\pi R^{2} \sigma T_{\rm eff}^{4}$, where $\sigma$ is the Stefan-Boltzmann constant and $L_{\gamma}$ is the thermal surface luminosity measured in the neutron star frame (see \cite{Ozel2013} for a detailed study of surface emission). For an arbitrary observer located at infinity - observer in the earth - the apparent luminosity is defined as $L_{\gamma}^{\infty} := L_{\gamma}\sqrt{1-2MG/Rc^{2}}$ resulting in the redshifted effective temperature $T^{\infty}_{\rm eff}=T_{\rm eff} (1+z)=T_{\rm eff} \sqrt{1-2MG/Rc^{2}}$, where $z$ is the redshift \cite{Yakovlev2004, Lattimer04}. In the Figure \ref{fig:teff} (see also Table \ref{tab:par} for specific values) we plotted the average of the effective temperature at infinity versus adiabatic index for models depicted in Figures  \ref{fig:neutronstars} and \ref{fig:fit}, the average was performed over all possible values of constant $K$ for EoS defined in equations \eqref{eq:polirest} (blue line) and \eqref{eq:politotal} (red line). Note that, for simplicity here we assume that $T_{\rm eff}=10^{6} \ K^{o}$ motivated in the temperatures estimated from pulsars \cite{Page2004, Ozel2013}. We found the redshifted effective temperatures in the range $6.3 \times 10^{5} K^{o}\ - \ 8.6 \times 10^{5}K^{o}$ for $p(\rho_{0})$ and $8.0 \times 10^{5} K^{o}\ - \ 8.4 \times 10^{5}K^{o}$ for $p(\rho)$, respectively.  We should remark that those models correspond to the stars with $C_s < c$, while for the maximum mass and radius values we get a reduced range $8.0 \ - \ 8.4 \times 10^{5} \ K^{o}$ in the two EoS models considered in this work.

\begin{table}
\tiny
\center
\label{tab:par}
\hspace{-0.3cm}
\begin{tabular}{ccccc} \hline
{\bf $\Gamma$} & $K$  & $T_{\rm eff}^{\infty}[K^{o}]$ & $\rho [\rho_{\rm{nuc}}]$ & $p {\rm [dyn/{cm}^{2}]}$ \\ \hline \hline
&$p=K\rho^{\Gamma}$\\
 \hline
{\bf $1.81$} & $30.5$       & $8.38\times 10^{5}$ & $8.60$ & $1.48 \times {10}^{29}$ \\
{\bf $1.87$} & $45-50.5$  & $8.46\times 10^{5}$& $8.31-9.50$ & $1.90-2.18 \times {10}^{30}$ \\ 
{\bf $1.93$} & $66-75$     & $8.15\times 10^{5}$& $8.94-10.24$ & $2.69-3.07 \times {10}^{30}$ \\ 
{\bf $2.02$} & $133$        & $8.00\times 10^{5}$& $9.84$  & $1.38 \times {10}^{33}$ \\ 
\hline \hline
& $p=K\rho_{0}^{\Gamma}$  \\  \hline
{\bf $2.05$} & $210$        & $8.37\times 10^{5}$ & $5.49$  & $1.90 \times {10}^{33}$ \\
{\bf $2.15$} & $398-465$ & $8.25\times 10^{5}$ & $5.70-6.54$  & $1.48-1.70 \times {10}^{35}$ \\ 
{\bf $2.24$} & $700-835$ & $8.15\times 10^{5}$ & $6.07-7.00$  & $7.02-8.10 \times {10}^{36}$ \\ 
{\bf $2.40$} & $2300$      & $8.02\times 10^{5}$& $6.68$ & $6.45 \times {10}^{39}$ \\  
\hline
\end{tabular}
\caption{Numerical outcomes for neutron stars modelled by polytropic EoS with adiabatic $\Gamma$ and adiabatic constant $K$ that satisfy the range of maximum masses and radii \cite{Rezzolla2017, Most2018}. In each approach, we report the effective temperature at infinity $T_{\rm eff \ \infty}$ in Kelvin degrees, pressure $p$ in cgs units ($\rm dyn \ / \ {cm}^{2}$) the corresponding energy $\rho$ and rest mass density $\rho_{0}$, respectively, in nuclear density units $\rho_{\rm {nuc}}$.}
\end{table}

\subsection{Quark stars} 

\begin{figure}[h!]
\center
\includegraphics[width=1.0\columnwidth]{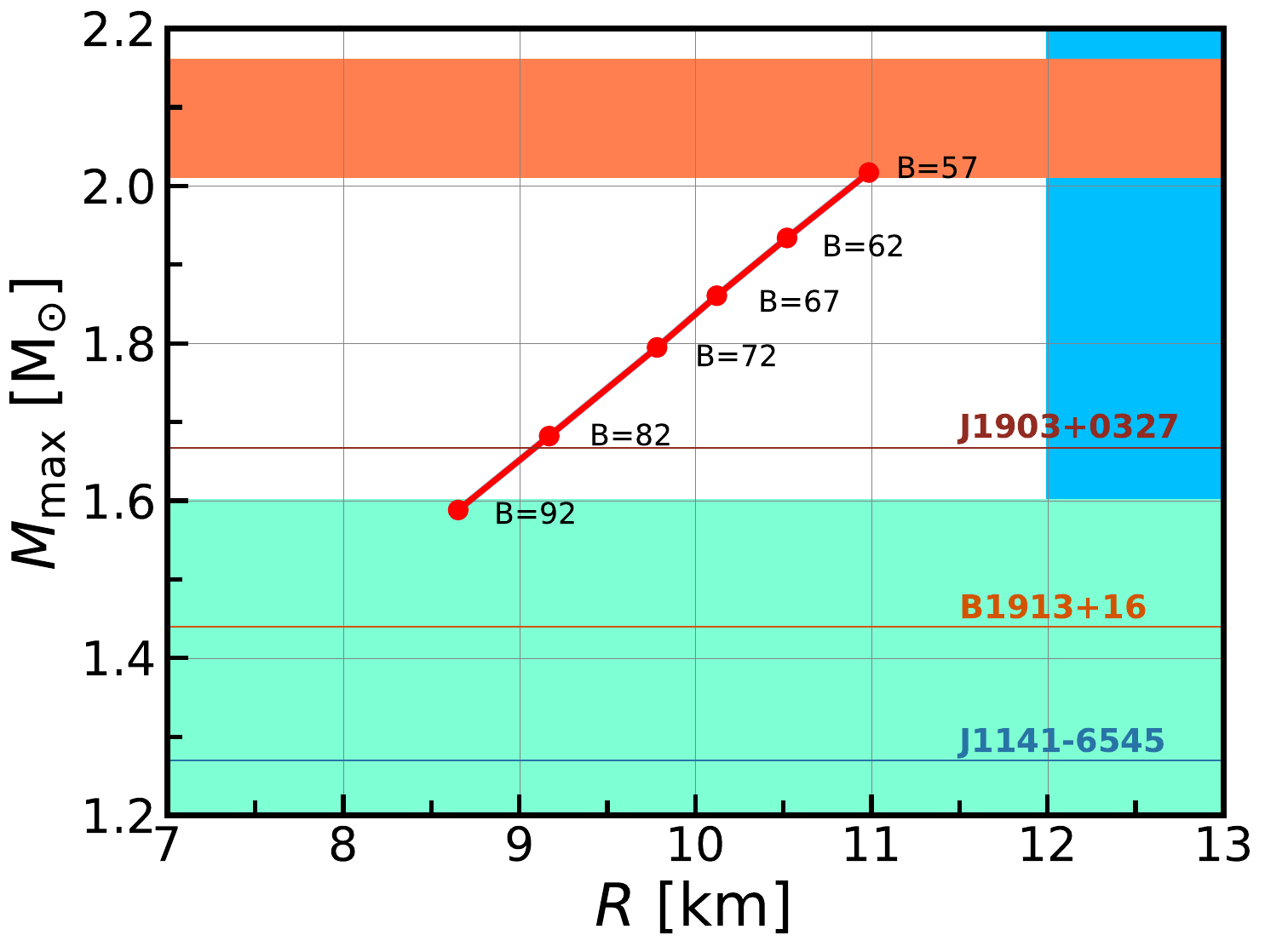}
\caption{Compactness for quark stars built following the MIT bag model. The dots correspond
to some particular vacuum energy in range $57 \ MeV/fm^{3} \ < \ B \ < \ 92\ MeV/fm^{3}$. In the same way, as in Figure \ref{fig:neutronstars} we include some representative masses estimated from pulsars, gravitational waves and maximum mass and radii constrictions.
\label{fig:MIT}}
\end{figure}

The numerical integration of TOV equations  (\ref{eq:mass}), (\ref{eq:press}), (\ref{eq:alp}) coupled 
to the constitutive relation (\ref{eq:MIT}) (the MIT bag-model EoS), describe a spherical symmetric 
non-rotating quark star. We summarize the maximum mass and its corresponding radius 
in Figure  \ref{fig:MIT}. 
Several simulations were carried out for different values of the vacuum energy density parameter, 
in particular, we explore values between the range $57~MeV/fm^{3}$ to $92 ~MeV/fm^{3}$. 
For the specific value of $B=60~MeV/fm^{3}$, we have obtained values of the mass and radius 
in agreement with the ones reported in \cite{gourgoulhon1999} for non-rotating quark stars. 
We have also found that quark stars modelled between the range of 
$90~MeV/fm^{3}<B<92~MeV/fm^{3}$ can reproduce the masses estimated in the event GW170817 
\cite{Abbott2017_etal}. Furthermore, to satisfy the constrained masses for compact 
stars reported in \cite{Rezzolla2017}, the values of $B$ must be in the range 
$49.5~MeV/fm^{3}<B<57.3~MeV/fm^{3}$; however, for these values,  we get more compact 
stars with radius in the range from $10.95$ $~{\rm km}$ to $11.78$ $~{\rm km}$, as is expected 
for quarks stars, which are more compact than neutron stars (see the blue shadowed region in 
Figure \ref{fig:MIT}, which corresponds to the acceptable radius of neutron stars 
\cite{Most2018}). Additionally to this fact, some of these energy values $B$ are out from the 
valid range and was excluded in the figure, obtaining that only one case satisfy the maximum 
mass constriction; $B=57~MeV/fm^{3}$ giving us a mass $M=2.1 \ M_{\odot}$ and $R=11 {\rm km}$. 
We found that the thermal emission from the surface of quark stars - following the same idea as 
in neutrons stars - give us an almost constant effective temperatures at infinity 
$T_{\rm eff}^{\infty}=7.95 \times 10^{5} \ K^{o}$ (see the Figure \ref{fig:teff}).


\subsection{Binding Energy} 

Using our numerical results we compute the binding energy for neutron and quark stars using the 
relation proposed in \cite{Lattimer01}
\begin{eqnarray}
\frac{BE}{M}=\frac{6 q}{5(2 - q)},
\label{eq:BE}
\end{eqnarray}
which is a function of the compactness $q=M/R$. This energy is relevant in the astrophysical 
context since $BE/M$ it is measured from the neutrinos emitted in a supernova explosion. 
We found that the binding energy are in the range $0.10427<BE/M<0.1194$ with compactness 
$0.159<q<0.181$  for neutron stars with total masses between $2.01<M<2.16 ~M_{\odot}$ 
and sizes of $12 < R < 13.45~{\rm km}$. For quark stars the binding energy obtained run in the 
range $0.12126<BE/M<0.12130$ and compactness $0.1835<q<0.1836$. The compactness of 
quark stars is greater than compactness in neutron stars, that means that same mass it is 
contained in a small region. The range of energies using a single layer polytropic and MIT EoS are 
in agreement with the values reported in \cite{Lattimer01}. 
 

\section{Summary}
\label{sec:Sum}

We have performed numerical simulations of non-rotating neutron and quark stars using single 
layer polytropic and MIT bag model equations of state, respectively. In particular, the polytropic 
EoS was applied to the rest mass and energy densities, keeping the models in which the sound 
speed is subluminal in the interior of the stars. In this first two cases for neutron stars, we found 
that the stars with adiabatic index in the range $1.81< \Gamma<2.02$ for EoS $p=K \rho^{\Gamma}$ 
and $2.05 < \Gamma<2.40$ for $p=K \rho_{0}^{\Gamma}$  are optimal to reproduce the 
constrained maximum masses and the corresponding radius recently reported by 
\cite{Rezzolla2017, Most2018}. We have carried out the same systematic search of parameters 
to cover the range of individual masses estimated from the GW170817 gravitational wave emission, 
the outcomes are showed in Figure \ref{fig:neutronstars} and values of $K$ reported 
in table \ref{tab:par}. Our numerical result for equation of state applied to the rest mass density 
are in agreement with recent {\it NICER} pulsar detection PSR J0030+0451  where a polytropic EoS 
with adiabatic index $\Gamma=2.5$ has been used  to describe the neutron star matter with mass 
$M=1.34_{-0.16}^{+0.15} \ M_{\odot}$ \cite{NICER2019L21, NICER2019L22}. 
 
On the other hand, for the case of quark stars, the constrained maximum masses computed in 
\cite{Rezzolla2017, Most2018} are reached for a single bag energy value $B=57 \ MeV/fm^{3}$ 
giving us a mass $M=2.1 \ M_{\odot}$ and radius $R=11 \ {\rm km}$, respectively. Now If we assume that the gravitational 
waves measured in GW170817 come from quarks stars, the parameter $B$ must be in the range 
$90 \ MeV/fm^{3} \ < \ B \ < \ 92$ $MeV/fm^{3}$ and their  respective radius are 
$\sim 8.65-8.75~{\rm km}$, which gives more compact stars.  However, our results for the range 
$57~MeV/fm^{3}$ $<B<92$ $~MeV/fm^{3}$ can also reproduce the recent estimated mass from 
gravitational wave detection GW190425 \citep{Abbott2019}. Notable differences have been found 
in the radius of neutron and quark stars, the last one gives us more compact stars, i. e., smaller radius.

Together with compactness, we have carried a simplified calculation of effective temperatures measured 
at infinity by assuming the the temperature at the surface of the star is $T_{\rm eff}=10^{6}$  based in the 
pulsar observations \cite{Lattimer04,Page2004}.  We found that the redshifted effective temperatures for 
neutron stars are in the range $6.3 \times 10^{5} K^{o}\ - \ 8.6 \times 10^{5}K^{o}$, furthermore for quark stars
we have been found a constant $T_{\rm eff}^{\infty}=7.95 \times 10^{5} \ K^{o}$.

Finally, we have estimated the binding energy for neutron stars with both single layer polytropic 
equations of state and found that this quantity ranges from $BE/M = 0.10427$ to $BE/M = 0.1194$.  
For the case of the stars constituted by quark matter, the binding energy is $BE/M \sim 0.121$. 
From our numerical results - using the simple EoS - is possible to differentiate by using the set of parameter 
studied here what kind of star we are observing, in the case that it can be measured.

The outcomes presented here can be improved in many ways. First introducing the spin to the stars; is well known that
the rotation of the stars gives us a different range of masses and radii. Second, by including a more realistic EoS with 
multiple layers and crusts. Finally, consider the possibility of mixed matter in the star. We plan to address these features in future works.

\vspace{1cm}

\acknowledgments
ACO gratefully acknowledges to CONACYT Postdoctoral Fellowship 291168 and 291258.  F.D.L-C was supported in part by VIE-UIS, under Grant No. 2493 and by COLCIENCIAS, Colombia, under Grant No. 8863. CC acknowledges partial support by CONACyT Grant CB-2012-177519-F.

\bibliographystyle{spr-mp-nameyear-cnd}
 \bibliography{biblio-u1}

\begin{thebibliography}{41}
\ifx \bisbn   \undefined \def \bisbn  #1{ISBN #1}\fi
\ifx \binits  \undefined \def \binits#1{#1} \fi
\ifx \bauthor  \undefined \def \bauthor#1{#1} \fi
\ifx \batitle  \undefined \def \batitle#1{#1} \fi
\ifx \bjtitle  \undefined \def \bjtitle#1{#1}\fi
\ifx \bvolume  \undefined \def \bvolume#1{\textbf{#1}}\fi
\ifx \byear  \undefined \def \byear#1{#1} \fi
\ifx \bissue  \undefined \def \bissue#1{#1} \fi
\ifx \bfpage  \undefined \def \bfpage#1{#1} \fi
\ifx \blpage  \undefined \def \blpage #1{#1} \fi
\ifx \burl  \undefined \def \burl#1{\textsf{#1}} \fi
\ifx \doiurl  \undefined \def \doiurl#1{\textsf{#1}} \fi
\ifx \betal  \undefined \def \betal{\textit{et al.}} \fi
\ifx \binstitute  \undefined \def \binstitute#1{#1} \fi
\ifx \binstitutionaled  \undefined \def \binstitutionaled#1{#1} \fi
\ifx \bctitle  \undefined \def \bctitle#1{#1} \fi
\ifx \beditor  \undefined \def \beditor#1{#1} \fi
\ifx \bpublisher  \undefined \def \bpublisher#1{#1} \fi
\ifx \bbtitle  \undefined \def \bbtitle#1{#1} \fi
\ifx \bedition  \undefined \def \bedition#1{#1} \fi
\ifx \bseriesno  \undefined \def \bseriesno#1{#1} \fi
\ifx \blocation  \undefined \def \blocation#1{#1} \fi
\ifx \bsertitle  \undefined \def \bsertitle#1{#1} \fi
\ifx \bsnm \undefined \def \bsnm#1{#1} \fi
\ifx \bsuffix \undefined \def \bsuffix#1{#1} \fi
\ifx \bparticle \undefined \def \bparticle#1{#1} \fi
\ifx \barticle \undefined \def \barticle#1{#1} \fi
\ifx \bconfdate \undefined \def \bconfdate #1{#1} \fi
\ifx \botherref \undefined \def \botherref #1{#1} \fi
\ifx \url \undefined \def \url#1{\textsf{#1}} \fi
\ifx \bchapter \undefined \def \bchapter#1{#1} \fi
\ifx \bbook \undefined \def \bbook#1{#1} \fi
\ifx \bcomment \undefined \def \bcomment#1{#1} \fi
\ifx \oauthor \undefined \def \oauthor#1{#1} \fi
\ifx \citeauthoryear \undefined \def \citeauthoryear#1{#1} \fi
\ifx \endbibitem  \undefined \def \endbibitem {}\fi
\ifx \bconflocation  \undefined \def \bconflocation#1{#1} \fi
\ifx \arxivurl  \undefined \def \arxivurl#1{\textsf{#1}} \fi

\bibitem[\protect\citeauthoryear{Abbott et~al.}{2017}]{Abbott2017_etal}
\begin{barticle}
\bauthor{\bsnm{Abbott}, \binits{B.P.}}, \betal:
\bjtitle{Phys. Rev. Lett.}
\bvolume{119},
\bfpage{161101}
(\byear{2017}).
doi:\doiurl{10.1103/PhysRevLett.119.161101}
\end{barticle}
\endbibitem

\bibitem[\protect\citeauthoryear{{Alcock} et~al.}{1986}]{Alcock86}
\begin{barticle}
\bauthor{\bsnm{{Alcock}}, \binits{C.}},
\bauthor{\bsnm{{Farhi}}, \binits{E.}},
\bauthor{\bsnm{{Olinto}}, \binits{A.}}:
\bjtitle{Astrophys. J.}
\bvolume{310},
\bfpage{261}
(\byear{1986}).
doi:\doiurl{10.1086/164679}
\end{barticle}
\endbibitem

\bibitem[\protect\citeauthoryear{{Andersson} et~al.}{2011}]{Andersson:2009yt}
\begin{barticle}
\bauthor{\bsnm{{Andersson}}, \binits{N.}},
\bauthor{\bsnm{{Ferrari}}, \binits{V.}},
\bauthor{\bsnm{{Jones}}, \binits{D.I.}},
\bauthor{\bsnm{{Kokkotas}}, \binits{K.D.}},
\bauthor{\bsnm{{Krishnan}}, \binits{B.}},
\bauthor{\bsnm{{Read}}, \binits{J.S.}},
\bauthor{\bsnm{{Rezzolla}}, \binits{L.}},
\bauthor{\bsnm{{Zink}}, \binits{B.}}:
\bjtitle{General Relativity and Gravitation}
\bvolume{43},
\bfpage{409}
(\byear{2011}).
\arxivurl{0912.0384}.
doi:\doiurl{10.1007/s10714-010-1059-4}
\end{barticle}
\endbibitem

\bibitem[\protect\citeauthoryear{Baiotti and Rezzolla}{2017}]{Baiotti2016}
\begin{barticle}
\bauthor{\bsnm{Baiotti}, \binits{L.}},
\bauthor{\bsnm{Rezzolla}, \binits{L.}}:
\bjtitle{Rept. Prog. Phys.}
\bvolume{80}(\bissue{9}),
\bfpage{096901}
(\byear{2017}).
\arxivurl{1607.03540}.
doi:\doiurl{10.1088/1361-6633/aa67bb}
\end{barticle}
\endbibitem

\bibitem[\protect\citeauthoryear{{Becerra-Vergara}
  et~al.}{2019}]{Becerra-Vergara:2019uzm}
\begin{barticle}
\bauthor{\bsnm{{Becerra-Vergara}}, \binits{E.A.}},
\bauthor{\bsnm{{Mojica}}, \binits{S.}},
\bauthor{\bsnm{{Lora-Clavijo}}, \binits{F.D.}},
\bauthor{\bsnm{{Cruz-Osorio}}, \binits{A.}}:
\bjtitle{\prd}
\bvolume{100}(\bissue{10}),
\bfpage{103006}
(\byear{2019}).
\arxivurl{1903.03047}.
doi:\doiurl{10.1103/PhysRevD.100.103006}
\end{barticle}
\endbibitem

\bibitem[\protect\citeauthoryear{{Bodmer}}{1971}]{Bodmer1971}
\begin{barticle}
\bauthor{\bsnm{{Bodmer}}, \binits{A.R.}}:
\bjtitle{Phys. Rev. D}
\bvolume{4},
\bfpage{1601}
(\byear{1971})
\end{barticle}
\endbibitem

\bibitem[\protect\citeauthoryear{{Chodos} et~al.}{1974}]{chodos1974}
\begin{barticle}
\bauthor{\bsnm{{Chodos}}, \binits{A.}},
\bauthor{\bsnm{{Jaffe}}, \binits{R.L.}},
\bauthor{\bsnm{{Johnson}}, \binits{K.}},
\bauthor{\bsnm{{Thorn}}, \binits{C.B.}},
\bauthor{\bsnm{{Weisskopf}}, \binits{V.F.}}:
\bjtitle{Phys. Rev. D}
\bvolume{9},
\bfpage{3471}
(\byear{1974})
\end{barticle}
\endbibitem

\bibitem[\protect\citeauthoryear{{Cruz-Osorio} and
  {Lora-Clavijo}}{2016}]{Cruz2016}
\begin{barticle}
\bauthor{\bsnm{{Cruz-Osorio}}, \binits{A.}},
\bauthor{\bsnm{{Lora-Clavijo}}, \binits{F.D.}}:
\bjtitle{Mon. Not. R. Astron. Soc.}
\bvolume{460},
\bfpage{3193}
(\byear{2016}).
doi:\doiurl{10.1093/mnras/stw1149}
\end{barticle}
\endbibitem

\bibitem[\protect\citeauthoryear{{Cruz-Osorio} et~al.}{2012}]{Cruz2012}
\begin{barticle}
\bauthor{\bsnm{{Cruz-Osorio}}, \binits{A.}},
\bauthor{\bsnm{{Lora-Clavijo}}, \binits{F.D.}},
\bauthor{\bsnm{{Guzm{\'a}n}}, \binits{F.S.}}:
\bjtitle{Mon. Not. R. Astron. Soc.}
\bvolume{426},
\bfpage{732}
(\byear{2012}).
doi:\doiurl{10.1111/j.1365-2966.2012.21794.x}
\end{barticle}
\endbibitem

\bibitem[\protect\citeauthoryear{Cruz-Osorio
  et~al.}{2017}]{Cruz-Osorio:2017epa}
\begin{barticle}
\bauthor{\bsnm{Cruz-Osorio}, \binits{A.}},
\bauthor{\bsnm{Sanchez-Salcedo}, \binits{F.J.}},
\bauthor{\bsnm{Lora-Clavijo}, \binits{F.D.}}:
\bjtitle{Mon. Not. Roy. Astron. Soc.}
\bvolume{471}(\bissue{3}),
\bfpage{3127}
(\byear{2017}).
\arxivurl{1707.05548}.
doi:\doiurl{10.1093/mnras/stx1815}
\end{barticle}
\endbibitem

\bibitem[\protect\citeauthoryear{{East} et~al.}{2016}]{East2016}
\begin{barticle}
\bauthor{\bsnm{{East}}, \binits{W.E.}},
\bauthor{\bsnm{{Paschalidis}}, \binits{V.}},
\bauthor{\bsnm{{Pretorius}}, \binits{F.}},
\bauthor{\bsnm{{Shapiro}}, \binits{S.L.}}:
\bjtitle{Phys. Rev. D}
\bvolume{93}(\bissue{2}),
\bfpage{024011}
(\byear{2016})
\end{barticle}
\endbibitem

\bibitem[\protect\citeauthoryear{{Farhi} and {Jaffe}}{1984}]{Farhi1984}
\begin{barticle}
\bauthor{\bsnm{{Farhi}}, \binits{E.}},
\bauthor{\bsnm{{Jaffe}}, \binits{R.L.}}:
\bjtitle{Phys. Rev. D}
\bvolume{30},
\bfpage{2379}
(\byear{1984})
\end{barticle}
\endbibitem

\bibitem[\protect\citeauthoryear{Flores et~al.}{2017}]{Flores:2017kte}
\begin{barticle}
\bauthor{\bsnm{Flores}, \binits{C.V.}},
\bauthor{\bsnm{Hall}, \binits{Z.B.} \bsuffix{II}},
\bauthor{\bsnm{Jaikumar}, \binits{P.}}:
\bjtitle{Phys. Rev.}
\bvolume{C96}(\bissue{6}),
\bfpage{065803}
(\byear{2017}).
doi:\doiurl{10.1103/PhysRevC.96.065803}
\end{barticle}
\endbibitem

\bibitem[\protect\citeauthoryear{{Gourgoulhon} et~al.}{1999}]{gourgoulhon1999}
\begin{barticle}
\bauthor{\bsnm{{Gourgoulhon}}, \binits{E.}},
\bauthor{\bsnm{{Haensel}}, \binits{P.}},
\bauthor{\bsnm{{Livine}}, \binits{R.}},
\bauthor{\bsnm{{Paluch}}, \binits{E.}},
\bauthor{\bsnm{{Bonazzola}}, \binits{S.}},
\bauthor{\bsnm{{Marck}}, \binits{J.-A.}}:
\bjtitle{Astron. Astrophys.}
\bvolume{349},
\bfpage{851}
(\byear{1999}).
\arxivurl{astro-ph/9907225}
\end{barticle}
\endbibitem

\bibitem[\protect\citeauthoryear{{Guzman} et~al.}{2012}]{Guzman2012}
\begin{botherref}
\oauthor{\bsnm{{Guzman}}, \binits{F.S.}},
\oauthor{\bsnm{{Lora-Clavijo}}, \binits{F.D.}},
\oauthor{\bsnm{{Morales}}, \binits{M.D.}}:
ArXiv e-prints
(2012).
\arxivurl{1212.1421}
\end{botherref}
\endbibitem

\bibitem[\protect\citeauthoryear{{Itoh}}{1970}]{Itoh70}
\begin{barticle}
\bauthor{\bsnm{{Itoh}}, \binits{N.}}:
\bjtitle{Progress of Theoretical Physics}
\bvolume{44},
\bfpage{291}
(\byear{1970})
\end{barticle}
\endbibitem

\bibitem[\protect\citeauthoryear{{Lai} et~al.}{2017}]{Lai2017b}
\begin{botherref}
\oauthor{\bsnm{{Lai}}, \binits{X.Y.}},
\oauthor{\bsnm{{Yu}}, \binits{Y.W.}},
\oauthor{\bsnm{{Zhou}}, \binits{E.P.}},
\oauthor{\bsnm{{Li}}, \binits{Y.Y.}},
\oauthor{\bsnm{{Xu}}, \binits{R.X.}}:
Research in Astronomy and Astrophysics,
(2017)
\end{botherref}
\endbibitem

\bibitem[\protect\citeauthoryear{{Lattimer} and {Prakash}}{2001}]{Lattimer01}
\begin{barticle}
\bauthor{\bsnm{{Lattimer}}, \binits{J.M.}},
\bauthor{\bsnm{{Prakash}}, \binits{M.}}:
\bjtitle{Astrophysical Journal}
\bvolume{550},
\bfpage{426}
(\byear{2001}).
\arxivurl{astro-ph/0002232}.
doi:\doiurl{10.1086/319702}
\end{barticle}
\endbibitem

\bibitem[\protect\citeauthoryear{{Lattimer} and {Prakash}}{2004}]{Lattimer04}
\begin{barticle}
\bauthor{\bsnm{{Lattimer}}, \binits{J.M.}},
\bauthor{\bsnm{{Prakash}}, \binits{M.}}:
\bjtitle{Science}
\bvolume{304},
\bfpage{536}
(\byear{2004}).
\arxivurl{arXiv:astro-ph/0405262}.
doi:\doiurl{10.1126/science.1090720}
\end{barticle}
\endbibitem

\bibitem[\protect\citeauthoryear{{Limousin} et~al.}{2005}]{limousin2005}
\begin{barticle}
\bauthor{\bsnm{{Limousin}}, \binits{F.}},
\bauthor{\bsnm{{Gondek-Rosi{\'n}ska}}, \binits{D.}},
\bauthor{\bsnm{{Gourgoulhon}}, \binits{E.}}:
\bjtitle{Phys. Rev. D}
\bvolume{71}(\bissue{6}),
\bfpage{064012}
(\byear{2005}).
doi:\doiurl{10.1103/PhysRevD.71.064012}
\end{barticle}
\endbibitem

\bibitem[\protect\citeauthoryear{{Lora-Clavijo} and
  {Guzm{\'a}n}}{2013}]{Lora2013}
\begin{barticle}
\bauthor{\bsnm{{Lora-Clavijo}}, \binits{F.D.}},
\bauthor{\bsnm{{Guzm{\'a}n}}, \binits{F.S.}}:
\bjtitle{\mnras}
\bvolume{429},
\bfpage{3144}
(\byear{2013}).
doi:\doiurl{10.1093/mnras/sts573}
\end{barticle}
\endbibitem

\bibitem[\protect\citeauthoryear{{Lora-Clavijo} et~al.}{2015a}]{Lora2015}
\begin{barticle}
\bauthor{\bsnm{{Lora-Clavijo}}, \binits{F.D.}},
\bauthor{\bsnm{{Cruz-Osorio}}, \binits{A.}},
\bauthor{\bsnm{{Guzm{\'a}n}}, \binits{F.S.}}:
\bjtitle{The Astrophysical Journal Supplement Series}
\bvolume{218},
\bfpage{24}
(\byear{2015}a).
\arxivurl{1408.5846}.
doi:\doiurl{10.1088/0067-0049/218/2/24}
\end{barticle}
\endbibitem

\bibitem[\protect\citeauthoryear{{Lora-Clavijo} et~al.}{2015b}]{Lora2015219}
\begin{barticle}
\bauthor{\bsnm{{Lora-Clavijo}}, \binits{F.D.}},
\bauthor{\bsnm{{Cruz-Osorio}}, \binits{A.}},
\bauthor{\bsnm{{Moreno M{\'e}ndez}}, \binits{E.}}:
\bjtitle{\apjs}
\bvolume{219},
\bfpage{30}
(\byear{2015}b).
\arxivurl{1506.08713}
\end{barticle}
\endbibitem

\bibitem[\protect\citeauthoryear{{Most} et~al.}{2018}]{Most2018}
\begin{botherref}
\oauthor{\bsnm{{Most}}, \binits{E.R.}},
\oauthor{\bsnm{{Weih}}, \binits{L.R.}},
\oauthor{\bsnm{{Rezzolla}}, \binits{L.}},
\oauthor{\bsnm{{Schaffner-Bielich}}, \binits{J.}}:
arxiv:1803.00549
(2018).
\arxivurl{1803.00549}
\end{botherref}
\endbibitem

\bibitem[\protect\citeauthoryear{{{\"O}zel}}{2013}]{Ozel2013}
\begin{barticle}
\bauthor{\bsnm{{{\"O}zel}}, \binits{F.}}:
\bjtitle{Reports on Progress in Physics}
\bvolume{76}(\bissue{1}),
\bfpage{016901}
(\byear{2013}).
\arxivurl{1210.0916}.
doi:\doiurl{10.1088/0034-4885/76/1/016901}
\end{barticle}
\endbibitem

\bibitem[\protect\citeauthoryear{{Page} et~al.}{2004}]{Page2004}
\begin{barticle}
\bauthor{\bsnm{{Page}}, \binits{D.}},
\bauthor{\bsnm{{Lattimer}}, \binits{J.M.}},
\bauthor{\bsnm{{Prakash}}, \binits{M.}},
\bauthor{\bsnm{{Steiner}}, \binits{A.W.}}:
\bjtitle{\apjs}
\bvolume{155}(\bissue{2}),
\bfpage{623}
(\byear{2004}).
\arxivurl{astro-ph/0403657}.
doi:\doiurl{10.1086/424844}
\end{barticle}
\endbibitem

\bibitem[\protect\citeauthoryear{{Paschalidis} and
  {Stergioulas}}{2017}]{Paschalidis2017b}
\begin{barticle}
\bauthor{\bsnm{{Paschalidis}}, \binits{V.}},
\bauthor{\bsnm{{Stergioulas}}, \binits{N.}}:
\bjtitle{Living Reviews in Relativity}
\bvolume{20},
\bfpage{7}
(\byear{2017}).
\arxivurl{1612.03050}.
doi:\doiurl{10.1007/s41114-017-0008-x}
\end{barticle}
\endbibitem

\bibitem[\protect\citeauthoryear{{Raaijmakers} et~al.}{2019}]{NICER2019L22}
\begin{barticle}
\bauthor{\bsnm{{Raaijmakers}}, \binits{G.}},
\bauthor{\bsnm{{Riley}}, \binits{T.E.}},
\bauthor{\bsnm{{Watts}}, \binits{A.L.}},
\bauthor{\bsnm{{Greif}}, \binits{S.K.}},
\bauthor{\bsnm{{Morsink}}, \binits{S.M.}},
\bauthor{\bsnm{{Hebeler}}, \binits{K.}},
\bauthor{\bsnm{{Schwenk}}, \binits{A.}},
\bauthor{\bsnm{{Hinderer}}, \binits{T.}},
\bauthor{\bsnm{{Nissanke}}, \binits{S.}},
\bauthor{\bsnm{{Guillot}}, \binits{S.}},
\bauthor{\bsnm{{Arzoumanian}}, \binits{Z.}},
\bauthor{\bsnm{{Bogdanov}}, \binits{S.}},
\bauthor{\bsnm{{Chakrabarty}}, \binits{D.}},
\bauthor{\bsnm{{Gendreau}}, \binits{K.C.}},
\bauthor{\bsnm{{Ho}}, \binits{W.C.G.}},
\bauthor{\bsnm{{Lattimer}}, \binits{J.M.}},
\bauthor{\bsnm{{Ludlam}}, \binits{R.M.}},
\bauthor{\bsnm{{Wolff}}, \binits{M.T.}}:
\bjtitle{\apjl}
\bvolume{887}(\bissue{1}),
\bfpage{22}
(\byear{2019}).
\arxivurl{1912.05703}.
doi:\doiurl{10.3847/2041-8213/ab451a}
\end{barticle}
\endbibitem

\bibitem[\protect\citeauthoryear{{Rezzolla} et~al.}{2018}]{Rezzolla2017}
\begin{barticle}
\bauthor{\bsnm{{Rezzolla}}, \binits{L.}},
\bauthor{\bsnm{{Most}}, \binits{E.R.}},
\bauthor{\bsnm{{Weih}}, \binits{L.R.}}:
\bjtitle{Astrophys. J. Lett.}
\bvolume{852},
\bfpage{25}
(\byear{2018}).
\arxivurl{1711.00314}.
doi:\doiurl{10.3847/2041-8213/aaa401}
\end{barticle}
\endbibitem

\bibitem[\protect\citeauthoryear{{Riley} et~al.}{2019}]{NICER2019L21}
\begin{barticle}
\bauthor{\bsnm{{Riley}}, \binits{T.E.}},
\bauthor{\bsnm{{Watts}}, \binits{A.L.}},
\bauthor{\bsnm{{Bogdanov}}, \binits{S.}},
\bauthor{\bsnm{{Ray}}, \binits{P.S.}},
\bauthor{\bsnm{{Ludlam}}, \binits{R.M.}},
\bauthor{\bsnm{{Guillot}}, \binits{S.}},
\bauthor{\bsnm{{Arzoumanian}}, \binits{Z.}},
\bauthor{\bsnm{{Baker}}, \binits{C.L.}},
\bauthor{\bsnm{{Bilous}}, \binits{A.V.}},
\bauthor{\bsnm{{Chakrabarty}}, \binits{D.}},
\bauthor{\bsnm{{Gendreau}}, \binits{K.C.}},
\bauthor{\bsnm{{Harding}}, \binits{A.K.}},
\bauthor{\bsnm{{Ho}}, \binits{W.C.G.}},
\bauthor{\bsnm{{Lattimer}}, \binits{J.M.}},
\bauthor{\bsnm{{Morsink}}, \binits{S.M.}},
\bauthor{\bsnm{{Strohmayer}}, \binits{T.E.}}:
\bjtitle{\apjl}
\bvolume{887}(\bissue{1}),
\bfpage{21}
(\byear{2019}).
\arxivurl{1912.05702}.
doi:\doiurl{10.3847/2041-8213/ab481c}
\end{barticle}
\endbibitem

\bibitem[\protect\citeauthoryear{{Schmitt}}{2010}]{Schmitt2010}
\begin{bbook}
\beditor{\bsnm{{Schmitt}}, \binits{A.}} (ed.):
\bbtitle{Dense Matter in Compact Stars}.
\bsertitle{Lecture Notes in Physics, Berlin Springer Verlag},
vol. \bseriesno{811}
(\byear{2010}).
\arxivurl{1001.3294}.
doi:\doiurl{10.1007/978-3-642-12866-0}
\end{bbook}
\endbibitem

\bibitem[\protect\citeauthoryear{Shu and Osher}{1988}]{Shu88}
\begin{barticle}
\bauthor{\bsnm{Shu}, \binits{C.W.}},
\bauthor{\bsnm{Osher}, \binits{S.J.}}:
\bjtitle{J. Comput. Phys.}
\bvolume{77},
\bfpage{439}
(\byear{1988})
\end{barticle}
\endbibitem

\bibitem[\protect\citeauthoryear{{The LIGO Scientific Collaboration}
  et~al.}{2017}]{Abbott2017b}
\begin{barticle}
\bauthor{\bsnm{{The LIGO Scientific Collaboration}}},
\bauthor{\bsnm{{the Virgo Collaboration}}},
\bauthor{\bsnm{{Abbott}}, \binits{B.P.}},
\bauthor{\bsnm{{Abbott}}, \binits{R.}},
\bauthor{\bsnm{{Abbott}}, \binits{T.D.}},
\bauthor{\bsnm{{Acernese}}, \binits{F.}},
\bauthor{\bsnm{{Ackley}}, \binits{K.}},
\bauthor{\bsnm{a{Adams}}, \binits{C.}},
\bauthor{\bsnm{{Adams}}, \binits{T.}},
\bauthor{\bsnm{{Addesso}}, \binits{P.}},
\bauthor{\bparticle{et} \bsnm{al.}}:
\bjtitle{Astrophys. J. Lett.}
\bvolume{848}(\bissue{2}),
\bfpage{12}
(\byear{2017})
\end{barticle}
\endbibitem

\bibitem[\protect\citeauthoryear{{The LIGO Scientific Collaboration}
  et~al.}{2020}]{Abbott2019}
\begin{botherref}
\oauthor{\bsnm{{The LIGO Scientific Collaboration}}},
\oauthor{\bsnm{{the Virgo Collaboration}}},
\oauthor{\bsnm{{Abbott}}, \binits{B.P.}},
\oauthor{\bsnm{{Abbott}}, \binits{R.}},
\oauthor{\bsnm{{Abbott}}, \binits{T.D.}},
\oauthor{\bsnm{{Abraham}}, \binits{S.}},
\oauthor{\bsnm{{Acernese}}, \binits{F.}},
\oauthor{\bsnm{{Ackley}}, \binits{K.}},
\oauthor{\bsnm{{Adams}}},
\oauthor{\bparticle{et} \bsnm{al}}:
arXiv e-prints,
2001
(2020).
\arxivurl{2001.01761}
\end{botherref}
\endbibitem

\bibitem[\protect\citeauthoryear{{Tooper}}{1964}]{Tooper64}
\begin{botherref}
\oauthor{\bsnm{{Tooper}}, \binits{R.F.}}:
{General Relativistic Polytropic Fluid Spheres.}
\textbf{140},
434
(1964).
doi:\doiurl{10.1086/147939}
\end{botherref}
\endbibitem

\bibitem[\protect\citeauthoryear{Tooper}{1965}]{tooper_1965_afs}
\begin{barticle}
\bauthor{\bsnm{Tooper}, \binits{R.F.}}:
\bjtitle{Astrophys. J.}
\bvolume{142},
\bfpage{1541}
(\byear{1965})
\end{barticle}
\endbibitem

\bibitem[\protect\citeauthoryear{{Tsokaros} et~al.}{2017}]{tsokaros2017}
\begin{barticle}
\bauthor{\bsnm{{Tsokaros}}, \binits{A.}},
\bauthor{\bsnm{{Ruiz}}, \binits{M.}},
\bauthor{\bsnm{{Paschalidis}}, \binits{V.}},
\bauthor{\bsnm{{Shapiro}}, \binits{S.L.}},
\bauthor{\bsnm{{Baiotti}}, \binits{L.}},
\bauthor{\bsnm{{Ury{\= u}}}, \binits{K.}}:
\bjtitle{Phys. Rev. D}
\bvolume{95}(\bissue{12}),
\bfpage{124057}
(\byear{2017}).
\arxivurl{1704.00038}.
doi:\doiurl{10.1103/PhysRevD.95.124057}
\end{barticle}
\endbibitem

\bibitem[\protect\citeauthoryear{Witten}{1984}]{Witten84}
\begin{barticle}
\bauthor{\bsnm{Witten}, \binits{E.}}:
\bjtitle{Phys. Rev. D}
\bvolume{30},
\bfpage{272}
(\byear{1984})
\end{barticle}
\endbibitem

\bibitem[\protect\citeauthoryear{{Yakovlev} et~al.}{2004}]{Yakovlev2004}
\begin{barticle}
\bauthor{\bsnm{{Yakovlev}}, \binits{D.G.}},
\bauthor{\bsnm{{Levenfish}}, \binits{K.P.}},
\bauthor{\bsnm{{Potekhin}}, \binits{A.Y.}},
\bauthor{\bsnm{{Gnedin}}, \binits{O.Y.}},
\bauthor{\bsnm{{Chabrier}}, \binits{G.}}:
\bjtitle{\aap}
\bvolume{417},
\bfpage{169}
(\byear{2004}).
\arxivurl{astro-ph/0310259}.
doi:\doiurl{10.1051/0004-6361:20034191}
\end{barticle}
\endbibitem

\bibitem[\protect\citeauthoryear{{Zhou} et~al.}{2018a}]{Zhou2018a}
\begin{barticle}
\bauthor{\bsnm{{Zhou}}, \binits{E.}},
\bauthor{\bsnm{{Tsokaros}}, \binits{A.}},
\bauthor{\bsnm{{Rezzolla}}, \binits{L.}},
\bauthor{\bsnm{{Xu}}, \binits{R.}},
\bauthor{\bsnm{{Ury{\= u}}}, \binits{K.}}:
\bjtitle{Universe}
\bvolume{4},
\bfpage{48}
(\byear{2018}a).
doi:\doiurl{10.3390/universe4030048}
\end{barticle}
\endbibitem

\bibitem[\protect\citeauthoryear{{Zhou} et~al.}{2018b}]{Zhou2018}
\begin{barticle}
\bauthor{\bsnm{{Zhou}}, \binits{E.}},
\bauthor{\bsnm{{Tsokaros}}, \binits{A.}},
\bauthor{\bsnm{{Rezzolla}}, \binits{L.}},
\bauthor{\bsnm{{Xu}}, \binits{R.}},
\bauthor{\bsnm{{Ury{\= u}}}, \binits{K.}}:
\bjtitle{\prd}
\bvolume{97}(\bissue{2}),
\bfpage{023013}
(\byear{2018}b).
doi:\doiurl{10.1103/PhysRevD.97.023013}
\end{barticle}
\endbibitem

\end{thebibliography}
\end{document}